\documentstyle[12pt,aaspp4,psfig]{article}

\def\msun{{\rm M}_\odot}

\begin{document}
 
\title{Non-LTE, Relativistic Accretion Disk Fits to 3C~273 and the Origin
of the Lyman Limit Spectral Break}
 
\author{Omer Blaes}
\affil{Department of Physics, University of California, Santa Barbara,
CA 93106}

\author{Ivan Hubeny}
\affil{AURA/NOAO, NASA Goddard Space Flight Center, Code 681, Greenbelt, 
       MD 20771}

\author{Eric Agol\footnote{Chandra fellow}~}
\affil{Theoretical Astrophysics, MS 130-33, California Institute of Technology,
           Pasadena, CA 91125}

\and
\author{Julian H. Krolik}
 
\affil{Department of Physics and Astronomy, Johns Hopkins University,
           Baltimore, MD 21218}
 
\begin{abstract}
We fit general relativistic, geometrically thin accretion disk models with
non-LTE atmospheres to near simultaneous multiwavelength data of 3C~273,
extending from the optical to the far ultraviolet.  Our model fits show no
flux discontinuity associated with a hydrogen Lyman edge, but they do exhibit
a spectral break which qualitatively resembles that seen in the data.  This
break arises from relativistic smearing of Lyman emission edges which are
produced locally at tens of gravitational radii in the disk.  We discuss the
possible effects of metal line blanketing on the model spectra, as well as the
substantial Comptonization required to explain the observed soft X-ray excess.
Our best fit accretion disk model underpredicts the near ultraviolet emission
in this source, and also has an optical spectrum which is too red.  We discuss
some of the remaining physical uncertainties, and suggest in particular that
an extension of our models to the slim disk regime and/or including nonzero
magnetic torques across the innermost stable circular orbit may help resolve
these discrepancies.
\end{abstract}


\section{Introduction}

Accretion disks around supermassive black holes are widely believed to be
the source of the optical/ultraviolet continuum in many classes of active
galactic nuclei (AGN).  The reasons for this include the expectation that
gravitational potential energy is efficiently converted into radiation in
disks and the prediction that the emergent spectrum peaks in the optical or
ultraviolet.  In addition, a disk provides a possible site for producing the
X-ray reflection features observed in Seyferts (Nandra \& Pounds 1994),
including a relativistically broadened iron K$\alpha$ line (Tanaka et al.
1995).

The agreement between accretion disk model predictions and observations of
optical/ultraviolet radiation from AGN is far from satisfactory, however
(see e.g. Koratkar \& Blaes 1999, Krolik 1999a, and Collin 2001 for recent
reviews).  Among the oft-cited problems is the absence of observed flux
discontinuities at the Lyman limit of hydrogen (Antonucci, Kinney, \& Ford
1989; Koratkar, Kinney, \& Bohlin 1992), in contrast to the large absorption
edges which were predicted to exist in early theoretical investigations of
accretion disk atmospheres (Kolykhalov \& Sunyaev 1984).  However, a number of
effects can weaken these predicted edges.  Rather high photospheric densities
in the early models were in part responsible for the large absorption edges
(Czerny \& Pojma\'nski 1990).  Later investigations showed that for
sufficiently high accretion rates, lower photospheric densities were produced,
which reduced
the edges or even drove them into emission in individual atmosphere models
(Coleman 1993, Shields \& Coleman 1994, Hubeny \& Hubeny 1997, Sincell \&
Krolik 1997).  Non-LTE effects also reduce discontinuities at the Lyman
limit (Sun \& Malkan 1989; Shields \& Coleman 1994; St\"orzer, Hauschildt,
\& Allard 1994; Hubeny \& Hubeny 1997).  Comptonization by a hot corona
can smear out Lyman edges (Czerny \& Zbyszewska 1991; Lee, Kriss, \& Davidsen
1992; Hsu \& Blaes 1998).  Finally, many authors have shown that relativistic
Doppler effects and gravitational redshifts can smear out an edge in the
integrated spectrum of a disk (Sun \& Malkan 1989, Laor \& Netzer 1989,
Laor 1992, Lee et al. 1992, Coleman 1993, Shields \& Coleman 1994, Wehrse
1997).

A small fraction of objects ($\sim10$~percent of those
investigated; Koratkar 1998; Koratkar \& Blaes 1999) exhibit
``partial Lyman edges'' in which there is no sharp flux discontinuity, but
there is a change in continuum slope near 912~\AA\/.  Examples
include the type 1 Seyfert Mrk~335 (Zheng et al. 1995) and the QSO PG~1630+377
(Koratkar et al. 1995).  A spectral break near the Lyman limit is also
observed in the {\it Hubble Space Telescope}/FOS composite quasar spectrum
(Zheng et al. 1997).

Recently, Kriss et al. (1999, hereafter KDZL) have presented {\it Hopkins
Ultraviolet Telescope} ({\it HUT}) data for 3C~273 which also appear to show
a spectral break near the Lyman limit.  More precisely, they observed
the source at two epochs (December 1990 and March 1995), and found that
single power-laws failed to provide adequate fits to the far ultraviolet
continuum in both epochs.  Broken power-laws, with breaks at $\sim900$~\AA\/
in the quasar rest frame,
provided better fits, although the improvement was substantial for the
first epoch only.

Appenzeller et al. (1998) have also presented far ultraviolet data on this
quasar taken with the {\it ORFEUS-II} mission.  Their raw, reddened spectrum
shows no obvious features near the intrinsic Lyman limit of the quasar, but
they did not present any actual fits to the continuum.  They noted
that for their observations to be consistent with longer wavelength
{\it International Ultraviolet Explorer} ({\it IUE})
data, a turnover near an observed wavelength of 1200~\AA\/ ($\simeq1040$~\AA\/
in the quasar rest frame) was required.  This is the longest wavelength in
the {\it ORFEUS-II} data, and the {\it IUE} comparison data was not
simultaneous, so the reality of this turnover is uncertain.

3C~273 was the first AGN to have its optical continuum compared to
accretion disk models (Shields 1978).  KDZL also applied crude accretion
disk models and found a reasonable fit at both epochs.
Their models consisted of a multitemperature blackbody disk
orbiting a Schwarzschild black hole, with an ad hoc absorption edge at the
Lyman limit, together with a spherical Comptonizing corona of unit Thomson
depth.  The temperature of the corona was constrained so as to fit the
soft X-ray spectrum which was observed simultaneously with {\it BBXRT}
and nearly simultaneously with {\it ROSAT} during the December 1990 {\it HUT}
observations.  This epoch also had nearly simultaneous optical (KPNO) and
{\it IUE} data.  While doing a good job in the ultraviolet and the
X-rays, KDZL acknowledged that their fits had a number of problems.  First,
they failed to produce
a break as sharp as that observed near the Lyman limit, i.e. their model
overpredicted the far ultraviolet.  In addition, it underpredicted the optical
portion of the spectrum.  In order to fit the far ultraviolet break as well as
they could, their models all had high disk inclinations of $\sim60^\circ$.
As they noted, the disk is more likely to be nearly face-on given that the
source has a superluminal jet.  Finally, their best fit model to the
December 1990 data had an Eddington ratio of 0.46, violating their assumption
of a geometrically thin disk.

We have recently computed an extensive grid of non-LTE accretion
disk models for a wide range of black hole masses and accretion rates,
in both Schwarzschild and Kerr spacetimes (Hubeny et al. 2000, hereafter
HABK).  In agreement with earlier studies, flux discontinuities at the
Lyman limit are generally not present for high luminosity disks.  Moreover,
we find that breaks in the continuum slope near the Lyman limit are common.
A superluminal source like 3C~273 provides a stringent test of the Lyman
edge problem because relativistic smearing is smallest in a face-on disk.
We have fit nearly face-on models to the 3C~273 data and find a qualitatively
reasonable fit to the observed spectral break, with no flux discontinuity,
even with no Comptonization.  We therefore propose that spectral breaks
near the Lyman limit in quasars may simply be due to relativistic smearing of
edges produced at different radii in the accretion disk.

This paper is organized as follows.  In \S~2 we describe the model fits
to the data.  Our model fits are far from perfect, and in \S~3 we explore
some of the physical effects that we have not included which could affect
the spectra in the vicinity of the Lyman limit.  Then in \S~4 we discuss
the implications of these fits and compare them to work by previous authors.
We also discuss how the broadband fits might be improved by extending our models
into the regime of slim disks, or by including magnetic torques across the
innermost stable circular orbit.  We summarize our conclusions in \S~5.

\section{Accretion Disk Model Fits}

\subsection{{\it HUT} Data}

The observations and analysis of the December 1990 {\it BBXRT}, {\it HUT},
{\it IUE}, and KPNO data are described in detail in KDZL.  We first concentrate
on model fits to the {\it HUT} data, which contain the Lyman edge region.
We will discuss later the broadband spectrum at both shorter and longer
wavelengths.

The HABK grid of Kerr ($a/M=0.998$) models covers nine black hole masses
from $1/8\times10^9$ to $32\times10^9$~$\msun$, each value a factor of two
larger than the next smaller mass.  For each mass, eleven models were
computed with accretion rates a power of two times one solar mass
per year, again spaced by factors of two.  The highest accretion rate for
each mass was chosen to make the overall luminosity approximately
0.3 times the Eddington limit.  Spectra were computed using our
relativistic transfer function (Agol 1997) for an observer at infinity
viewing the disk at a discrete set of inclination angles $i$ given by
$\cos i=0.01$, 0.2, 0.4, 0.5, 0.6, 0.8, and 0.99.  The only remaining
quantity is the viscosity parameter $\alpha$.  HABK considered two values,
0.01 and 0.1, but found that this made little difference to the resulting
spectra except at very high frequencies in the He~II Lyman continuum.
As this lies in the unobserved gap between the far ultraviolet and soft X-rays
in 3C~273, we consider only the $\alpha=0.01$ models in this paper.

In our fitting process, we first transform our models from the quasar rest
frame to the observer frame.  We adopt a redshift of 0.158 for 3C~273, a Hubble
constant $H_0=70$~km~s$^{-1}$~Mpc$^{-1}$, and a spatially flat cosmology with
$\Omega_{\rm M}=1/3$ and $\Omega_{\Lambda}=2/3$.
We then accounted for Galactic extinction by reddening our model spectrum.
We considered values of $E(B-V)$ ranging from 0.02 (as determined from the
{\it DIRBE} dust infrared emission map of Schlegel, Finkbeiner, \& Davis 1998)
to 0.032 (as determined from the HI column density measurements of Lockman and
Savage 1995).\footnote{Our treatment differs slightly from KDZL, who assumed
a cosmology with zero deceleration parameter and
$H_0=75$~km~s$^{-1}$~Mpc$^{-1}$.  Our rest frame flux densities are therefore 23
percent higher than theirs.  In addition, KDZL adopted a Galactic
$E(B-V)$ of 0.032 in their paper, although their figures 6-8 in fact used
$E(B-V)=0.02$ (G. Kriss 2000, private communication).}~  Assuming $R_V=3.1$,
we then reddened our models using the Cardelli, Clayton, \& Mathis (1989)
extinction curve.  Unfortunately, the hydrogen Lyman limit of 3C~273 lies at
1056\AA\/, where the extinction curve is somewhat uncertain.  We comment on
this further in section 4 below.

At this point, we interpolated our model spectra onto the observed {\it HUT}
data wavelengths and then added quasar emission and Galactic absorption lines
to our models according to the broken power law continuum fits presented in
tables 4 and 6 of KDZL.  Following KDZL, we also accounted for the fifty lowest
energy Galactic hydrogen Lyman absorption lines as well as the Lyman continuum,
assuming a neutral hydrogen column density of $1.8\times10^{20}$~cm$^{-2}$, and
Voigt profiles with Doppler parameters of 15~km~s$^{-1}$ convolved with the
Gaussian instrument resolution profile of 3\AA\/ FWHM.

Finally, we measured $\chi^2$ by comparing our model with the observed
{\it HUT} data.  We excluded the observed wavelength regions 980-1000\AA\/,
1194-1238\AA\/, 1287-1319\AA\/, and 1820-1870\AA\/ from our fits because of
contamination by terrestrial airglow lines.  We also excluded all observed
wavelengths shortward of 915\AA\/ because these wavelengths are affected by
the highest order Lyman lines, which we did not model.

We performed this procedure for our entire grid of non-LTE models,
searching for the model with the mass $M$, accretion rate $\dot M$, and
inclination $i$ which had the lowest $\chi^2$.  Note that these fits were
made with the absolute spectral flux density, i.e. we fit both the shape
and normalization of the spectrum.  The best fit models turned out to be
$M=1.6\times10^{10}$~$\msun$, $\dot M=4$~$\msun$~yr$^{-1}$, and $\cos i=0.4$
and 0.5 for $E(B-V)=0.02$ and 0.032, respectively, with corresponding values
of $\chi^2=2522$ and 2866 for 1577 data points.  However, a number of models
turned out to have comparable values of $\chi^2$.  Due perhaps to the coarse
graining of accretion rates in our grid (successive values being separated by
factors of two), almost all these models had accretion rates of
4~$\msun$~yr$^{-1}$.  This was independent of the Galactic
extinction, and is clearly driven by the normalization of the spectrum.

Figure 1 illustrates how one of these models compares to the {\it HUT} data.
While the fit appears quite reasonable at long wavelengths, it does not do
a good job at the shortest wavelengths and in fact is still not adequate given
the value of $\chi^2$.  Part of the reason for the high value of $\chi^2$ may
be because we did not try refitting the lines with our continuum models.
As these lines are not the focus of this paper, we instead use this particular
fit to isolate six continuum windows at quasar rest wavelengths of 826-830,
864-872, 955-959, 1093-1099, 1354-1360, and 1445-1455\AA\/.\footnote{Recent
high resolution {\it Far Ultraviolet Spectroscopic Explorer} spectra of
3C~273 by Sembach et al. (2001) cover the first three wavelength regions and
confirm that they are devoid of large equivalent width absorption lines.}~
We performed
a weighted average of the data in these six regions, and then repeated
our entire fitting procedure, this time neglecting lines entirely, and now
letting the reddening float between 0.2 and 0.032.  The best fit model
was the same: $M=1.6\times10^{10}$~$\msun$, $\dot M=4$~$\msun$~yr$^{-1}$,
and $\cos i=0.4$ with $\chi^2=12$ for $E(B-V)=0.024$.
Given that we have only six data points, this is not a good fit statistically.
Figure 2 shows the reason: this particular model is unable to reproduce the
curvature in the turnover at the shortest wavelengths.

3C~273 is a superluminal source, so models with such high inclinations are
physically ruled out.  The observed VLBI proper
motions give apparent transverse velocities ranging from 7.0 to 11~$c$ for
our Hubble constant (e.g. Abraham et al. 1996).  The larger value directly
implies $\cos i> 0.98$.  However, the apparent velocities of the
superluminal components which originated near 1990, the year of the
multiwavelength campaign data considered here, were closer to the smaller
value, giving $\cos i>0.96$.  Abraham \& Romero (1999) have developed a
kinematic precession model of the 3C~273 jet which gives $\cos i\simeq0.98$
for the December 1990 epoch.

Again fitting to just the
continuum points, the best fit nearly face-on model ($\cos i=0.99$) was
$M=2\times10^{9}$~$\msun$, $\dot M=4$~$\msun$~yr$^{-1}$ and
$E(B-V)=0.0253$, giving $\chi^2=76$.  Figure 3 illustrates this model fit.
The model clearly does a better job at reproducing the short wavelength slope
(and therefore the overall spectral break) than the more edge-on model shown
in figure 2, but it also exhibits a bump that peaks redward of the quasar
rest-frame Lyman limit.  The origin of these features can be seen in figure
4, which shows a crude estimate of the contributions that individual annuli
make to the far ultraviolet region of the spectrum.  We multiplied the
local rest-frame emergent spectral flux from each annulus by an ``area''
computed in a Newtonian fashion from the Boyer-Lindquist coordinate radius
and width (figure 4a), and then added this power to that from other annuli
in the model (figure 4b).
This provides a quick illustration of what determines the total spectrum,
but we stress that our full model properly takes into account the general
relativistic proper area of each emitting element, as well as gravitational
redshifts, Doppler shifts, and bending of photon trajectories to the observer.
Figure 4 shows clearly that in the absence of these relativistic effects, the
disk would produce a Lyman edge in emission.  This edge is completely smeared
out, however, in models with significant inclination with respect to the
observer, such as in figure 2.  Gravitational redshifts and the transverse
Doppler effect also smear and redden this edge in the nearly face-on model
shown in figure 3, producing a bump redward of the rest-frame Lyman limit.
We stress that even in this case, there is no flux discontinuity, although
there is a marked change in spectral slope. 

HABK also computed a grid of models around Schwarzschild holes, and these
gave similar results.  The best fit non-LTE model to the continuum data
points was $M=4\times10^9$~$\msun$, $\dot M=22$~$\msun$~yr$^{-1}$,
$\cos i=0.5$, and $E(B-V)=0.032$, with $\chi^2=23$.  This model has the same
overall luminosity as the best fit Kerr models.  The best nearly face-on
Schwarzschild model had $M=10^9$~$\msun$, $\dot M=11$~$\msun$~yr$^{-1}$,
and $E(B-V)=0.032$, with $\chi^2=389$.  The Schwarzschild models are clearly
much worse than the Kerr models, suggesting that some black hole spin may be
required in this source.  This may confirm theoretical ideas that
jet production is related to black hole spin, but we caution the reader
not to draw strong conclusions given the poor quality of our fits.

\subsection{Optical/Ultraviolet Multiwavelength Data}

The December 1990 observing campaign also had near simultaneous {\it IUE}
and optical (KPNO) data.  The accretion disk is putatively responsible for
much of this emission as well, so we also tried fitting this data.  In terms
of rest wavelengths, we identified 1795-1805\AA\/ as a continuum window in the
{\it IUE}/LWP data,  and 4190-4210, 4700-4720, 5090-5110, and 5590-5610\AA\/
as continuum windows in the optical KPNO data.  Based on the emission line
studies of I~Zw~1 by Boroson \& Green (1992) and Laor et al. (1997), these
windows avoid most of the potential Fe~II line blends.  Unfortunately,
we did not have
access to the error bars on this data, so we estimated the variance by
calculating it directly using an unweighted average of the data points in
each wave band.  Combined with the six {\it HUT} continuum points, we therefore
had eleven points in total in our continuum dataset.

3C~273 is known to exhibit correlated variability between the optical
and infrared (Cutri et al. 1985), suggesting some contamination
of the big blue bump by emission from the jet.
However, the median optical polarization is less than 0.5~percent
(Impey, Malkan \& Tapia 1989),
and the polarized flux can be fit with a steep power law
$F_\nu\propto\nu^{-1.9}$ (Wills 1989).  Jet contamination
is therefore negligible throughout the far ultraviolet {\it HUT} data,
but it might be an important contributor in the optical.
We therefore considered models with an additional $F_\nu\propto\nu^{-1.9}$
power law, normalized to be ten percent of the observed optical
(5500~\AA\/) emission (Impey et al. 1989).\footnote{Our fits to the {\it HUT}
data alone which we discussed previously also included this miniblazar
component.  Because it is so small in the far ultraviolet, the best fit models
do not change, but the actual values of $\chi^2$ which we quoted do depend
slightly on this.}~

The best fit non-LTE, nearly face-on ($\cos i=0.99$) Kerr disk model to the
multiwavelength data was the same as that for the {\it HUT} data alone:
$M=2\times10^{9}$~$\msun$ and $\dot M=4$~$\msun$~yr$^{-1}$.  The best fit
extinction was $E(B-V)=0.0261$ with the blazar component ($\chi^2=187$)
and $E(B-V)=0.024$ without the blazar component ($\chi^2=126$).  Figures
5 and 6 illustrate these model fits to the data.

The slope of the optical continuum data is substantially bluer than our
disk model, a fact which is true for all the models we present in this
paper.  Fitting a power law through the four optical data points
gives a spectrum $F_\lambda\propto\lambda^{-2.43\pm0.12}$, or
$F_\nu\propto\nu^{0.43\pm0.12}$.  While this is bluer than most bright
quasars (e.g. Francis et al.  1991), we note that the reddened optical/near
infrared data presented in Lichti et al.  (1995) for this source are also
comparably blue.  Even if we applied no
reddening and no red blazar component to our best fit disk model, it would
only have an optical spectrum as blue as $F_{\nu}\propto\nu^0$.  It could be
that our optical continuum points suffer from contamination by emission lines.
Figure 6 shows clear evidence for broad line emission at the shortest optical
wavelengths, though as stated above we tried to avoid this in our choice
of continuum points.

\section{Modifications to the Basic Disk Models}

The HABK models leave out some potentially important physics.  Here we
consider two additional effects which may alter the spectrum of 3C~273
around the Lyman limit of hydrogen.

\subsection{Metal Line Blanketing}

The HABK models completely neglect line opacities from all atomic species.
Numerous ultraviolet lines of various metal species exist in
hot atmospheres, and this may significantly affect the emergent
spectrum of a disk in the Lyman limit region (Hubeny \& Hubeny 1998).

To gauge how important this is, we have recomputed the emergent spectrum
through our best fit Kerr disk model ($M=2\times10^{9}$~$\msun$ and
$\dot M=4$~$\msun$~yr$^{-1}$), including line opacities from the thirty
lightest elements (H-Zn).  We did this by recomputing the radiative transfer
in our converged atmosphere model for each annulus with line opacities included,
but we did not recompute the atmosphere model.  The line opacities and source
functions were taken to be in LTE, with thermal Doppler line profiles.
The resulting spectra
do not conserve flux because energy balance with the additional opacities
is not enforced.  The normalization of the spectrum is therefore not to
be trusted, but spectral features associated with the opacity sources may
survive a self-consistent treatment.  In any case, these spectra provide
an indication of the importance of metal line blanketing.

Figure 7 shows the resulting far ultraviolet spectrum, compared to that
without line blanketing.  The parameters are exactly the same as those of
our best fit model shown in Figure 3.  The qualitative features of the model
spectrum remain the same: there is still no flux discontinuity at the Lyman
limit, and there is still a spectral break.  However, line blanketing appears
to {\it enhance} the residual bump left over from smearing the Lyman emission
edges of individual annuli, by absorbing flux both longward and shortward of
the edge.  Whether this will be preserved in more self consistent modeling
which maintains flux conservation and includes non-LTE effects remains to be
seen.

\subsection{Comptonization and the Soft X-ray Excess of 3C~273}

We have recently computed a series of disk models which include the effects
of Comptonization within the disk structure itself, assuming that the energy
dissipation rate per unit mass is vertically constant within each annulus
(Hubeny et al. 2001).  For the black hole masses and accretion rates we have
found to be relevant for 3C~273, these Comptonized models differ negligibly
in the optical-ultraviolet region from the models computed by HABK which we
have used here.

The December 1990 observing campaign also included {\it BBXRT} and {\it ROSAT}
observations, which together show the presence of a clear soft X-ray excess.
Such emission is not present in any of our models (including the Comptonized
models of Hubeny et al. 2001).  A possible explanation for the soft excess
is the presence of an additional, much stronger, Comptonizing component
(e.g. KDZL).  Such a component might change the observed shape of the far
ultraviolet continuum.

In order to investigate this, we followed KDZL and Comptonized our disk spectra using the crude prescription of Czerny \& Zbyszewska (1991).  We treated the
disk spectrum (already folded through the relativistic transfer function) as
a point source in a spherical cloud of hot electrons described by two
parameters: an electron temperature and a radial Thomson depth.  It is likely
that the hot electrons are closer to the disk, and a better physical treatment
would be to locally Comptonize the spectra of each of our annuli, but our
simple-minded treatment should still be sufficient to capture the main
qualitative effects.

We defined soft X-ray fluxes by averaging the {\it ROSAT} data, corrected
for interstellar absorption, in quasar rest frame wavelength ranges of
30-40 and 40-50\AA\/.  We found an acceptable fit to these fluxes by
Comptonizing
our $M=2\times10^9$~M$_\odot$, $\dot M=4$~M$_\odot$~yr$^{-1}$, $\cos i=0.99$,
Kerr model with a cloud of unit radial Thomson depth and temperature
$3.3\times10^8$~K.  A reddening of $E(B-V)=0.032$ then provided the best fit
to the optical/ultraviolet data.
Comparisons between the Comptonized models and the data are shown in figures
8 and 9.

This Comptonized model provided an improved (though still
unacceptable) fit to the {\it HUT} data alone, giving $\chi^2\simeq50$.
Figure 9 shows why: Comptonization has smeared out the small Lyman edge
bump feature in the spectrum.
The overall quality of the fit to the optical and ultraviolet data remained
about the same, however, because Comptonization has little effect at longer
wavelengths.

\section{Discussion}

Quantitatively, all our model fits to the continuum data are rather poor.
There are numerous reasons why this might be so, as we will discuss later
in this section.  However, we still believe that our fits represent
a qualitative success, as there is no sign of a Lyman edge flux discontinuity.
This is true even though we are using face-on models, which minimize the
effects of relativistic Doppler smearing of the intrinsic Lyman edges formed
locally at each radius of the disk.  All that remains of the Lyman edge is a
small bump shifted redward from the Lyman limit.  As we showed in section 3,
this bump itself may well
be modified by the effects of line opacities and the Comptonization necessary
to produce the observed soft X-ray excess.  Although there is no Lyman edge,
there is a spectral break in the vicinity of the Lyman limit, in qualitative
agreement with that seen in the data.  In addition, while
our models do a poor job of explaining the slope of the optical data, they
are much better here than the model fits of KDZL which substantially
underpredicted the optical flux (cf. Fig. 7 of their paper).

Compared to the (Schwarzschild) fits of KDZL, our black hole mass is 2-3
times larger and our accretion rate is 3 times smaller.  Our best fit model
has a luminosity $\simeq0.3$ times the Eddington limit, i.e. is marginally
consistent with the disk being geometrically thin.  (This best-fit model
is in fact at the edge of our grid of models for this reason.)
Previous authors have obtained roughly similar values for the
black hole mass and accretion rate when trying to fit the overall shape
of the big blue bump with much simpler accretion disk models.  For
example, Shields (1978) found $M=10^9$~$\msun$ and $\dot M=3$~$\msun$~yr$^{-1}$
for a Newtonian multitemperature blackbody disk with an underlying power
law component extended from the infrared.  Ramos et al. (1997) fit
optical/ultraviolet
data taken in 1994 and 1995 with multitemperature modified blackbody disk
models around a maximal Kerr hole and found $M=10^9$~$\msun$ and an accretion
rate $\dot M=2.4$~$\msun$~yr$^{-1}$.  Due to the high radiative efficiency of
accretion onto a Kerr hole, this corresponds to a luminosity $\simeq0.5$ the
Eddington limit, too high for the disk to be truly geometrically thin.

Our accretion disk models also have rather large absorption edges near the
hydrogen Balmer limit.  As noted by HABK, these edges are not to be trusted
as they arise from cool annuli in the disk which have vertical density
inversions and steep, convectively unstable temperature profiles, at least
within our treatment of the vertical disk structure.  Direct comparison
of this region of the spectrum with the data is difficult due to contamination
by the ``little blue bump'', thought to be caused by Balmer continuum and
FeII line emission from the broad line region (BLR) of the quasar.

The annuli responsible for the emission near the Lyman limit are
also convectively unstable due to our assumed vertical dissipation profile
and the fact that they are supported by radiation pressure (Agol et al. 2001,
see also the discussion of Hubeny et al. 2001).  How this might affect the
Lyman edge locally at each annulus is unclear, and we are currently pursuing
models that incorporate vertical convective transport of heat, as well as
investigating different vertical profiles of the energy dissipation rate
(e.g. Miller \& Stone 2000).  However, radiation pressure may have even more
serious effects on the structure of the disk.  Turbulence in a radiation
dominated environment may result in strong density contrasts, invalidating our
approximation of a smooth density profile which depends only on height within
the annulus (Begelman 2001; Turner, Stone, \& Sano 2001).  Thermal and viscous
instabilities may also exist which might give rise to an inhomogeneous disk
structure (Krolik 1998), as well as invalidating the assumption of a stationary
flow.  Unfortunately, we cannot as yet predict how these effects might change
the results of our spectral computations.

Our current models also neglect reprocessing of external radiation, including
that due to bending of light rays from other parts of the disk.  This could be
important at far ultraviolet wavelengths, particularly in the near-maximal
Kerr models we showed here (Cunningham 1976).

We remind the reader that we tried fitting models corresponding to only two
black hole spins: $a/M=0$ and 0.998.  It could be that intermediate black
hole spins might produce improved fits.  To investigate this, we tried fitting
near face-on, multitemperature blackbody models to the data using a range of
accretion rates, black hole masses, and spins.  Figure 10 illustrates such
a model, corresponding to $a/M=0.7$.  This model produces a much better fit
to the far ultraviolet data ($\chi^2\simeq8$) than our best fit near
face-on disk model made from stellar atmospheres, although it is still
a statistically poor fit.  We stress that it contains little physics!
The smooth turnover at short wavelengths in the model spectrum is largely
produced by the Galactic reddening.  The fact that the far
ultraviolet fit is still poor suggests that a sharper spectral break than
can be produced by reddening alone is required.
Moreover, this model still fails to produce an optical spectrum as blue as
that observed.  We computed a hydrogen/helium disk model with stellar
atmospheres along the lines of the HABK models, with the same parameters as
this multitemperature blackbody model, and found a much worse fit than our
nearly maximal Kerr models.  We therefore believe that models with intermediate
black hole spins will not solve the remaining qualitative discrepancies in
our spectral fits.

A recent measurement of Galactic far ultraviolet extinction by Sasseen et
al. (2001) finds that the wavelength dependence is slightly flatter in shape
than the extrapolation of the Cardelli et al. (1989) curve we have used
throughout this paper.  This provides increased confidence that the observed
spectral break is in fact intrinsic to the quasar.  However, it is clearly
desirable to try fitting a miniblazar like 3C~273 which is at somewhat
higher redshift so that one could be certain that any spectral break
observed near the Lyman limit is less affected by the shape of the extinction
curve.

\subsection{Could the Accretion Flow in 3C~273 be a Slim Disk?}

Our models do a good job of explaining the lack of a Lyman edge in this
quasar, and in producing a far ultraviolet spectral break comparable to
what is observed.  However, they are far from producing a statistically
acceptable fit.  The chief discrepancies are a redder optical spectrum
than is observed, and an underprediction of the
$\lambda_{\rm obs}=1500-2000$\AA\/ ultraviolet emission.  In order to
resolve these discrepancies, one must alter the radial effective temperature
profile of the optically emitting annuli to be closer to the asymptotic
$r^{-3/4}$ law.  This will then
produce a bluer optical spectrum (as blue as $F_\nu\propto\nu^{1/3}$),
and would also therefore help enhance the $1500-2000$\AA\/ ultraviolet
emission.  At the same time, one must ensure that the overall luminosity of the
annuli responsible for the far ultraviolet emission remains approximately
the same, in order to continue to fit the observed spectrum in the region
of the Lyman limit.

It is noteworthy that our best fit model is already at the maximum Eddington
ratio $L/L_{\rm Edd}=0.3$ consistent with a geometrically thin disk.  Increasing
the Eddington ratio still further, either by lowering the black hole mass or
by increasing the accretion rate, would push the optically emitting annuli
further out in radius, where the $r^{-3/4}$ effective temperature profile
holds.  It would also move the disk into the regime of slim
disks (Szuszkiewicz, Malkan \& Abramowicz 1996).  Slim disks differ from
standard thin disks primarily in the role played by advection.  This flattens
the inner radial profile of effective temperature, thereby reducing the
far ultraviolet emission compared to what would be expected if the flow was
describable as a thin disk.  It also produces more extreme ultraviolet/soft
X-ray emission, an effect which will be enhanced by Comptonization within
the flow.

Another effect which could alter the models in the same direction is a
nonzero torque on the disk at the innermost stable circular
orbit.  Such a torque might be exerted through magnetic fields anchored in
the flow (Krolik 1999b, Gammie 1999).
Its primary effect is to enhance the radiative efficiency, which could be
compensated for by reducing the accretion rate.  At the same time, reprocessing
of the relatively enhanced inner disk radiation
at larger radii would cause the effective temperature profile to approach
$r^{-3/4}$ at smaller radii (Agol \& Krolik 2000), which would produce an
optical spectrum that is more blue.

It is possible that these effects will provide a much improved fit
to the 3C~273 data. We therefore intend to extend our detailed atmosphere
modeling to the slim disk regime in future, as well as investigate nonzero
inner torques.  We stress again that most quasars have optical
colors which are much redder than those of 3C~273, and therefore more in
agreement with the HABK disk models.  A parameter which distinguishes 3C~273
from other quasars, such as a high Eddington ratio, is therefore an attractive
possibility for explaining its unusually blue optical spectrum.

\section{Conclusions}

We have fit near face-on, bare accretion disk models to the multiwavelength
data obtained by KDZL in the December 1990 observing campaign of 3C~273.
These models
fully account for non-LTE effects in hydrogen and helium in the disk
atmospheres, and also include a full treatment of general relativistic
effects in the disk structure as well as photon propagation to the observer.
Our model fits are far from perfect, underpredicting the near ultraviolet
emission and producing an optical spectrum which is too red.  However, they
produce no flux discontinuities near the Lyman limit.  Instead, a small
bump in the spectrum exists, which arises from
gravitational redshifts and the transverse Doppler effect acting on individual
Lyman emission edges emitted locally at tens of gravitational radii.  We
performed a preliminary investigation of the effects of line blanketing, which
appear to enhance this bump, while the Comptonization required to explain
the observed soft X-ray excess smears it out.  Our models produce a spectral
break in the vicinity of the Lyman limit  which is in qualitative agreement
with observation.

There are a number of other problems with accretion disk models that we
have not addressed in this paper.  In particular, the near simultaneous
variability observed throughout the optical/ultraviolet in AGN
almost certainly requires that reprocessing contribute to the
emission at some level, and we have not incorporated such reprocessing
at all in our disk models.  Moreover,
some quasars with partial Lyman edges have dramatically increased polarization
blueward of the Lyman limit (Impey et al. 1995, Koratkar et al. 1995),
a problem which still has no satisfactory explanation within the accretion
disk paradigm (see Shields, Agol, \& Blaes 2000 for a recent review).

We nevertheless believe that the model fits shown here demonstrate
substantially improved agreement between theory and observation of the
Lyman limit region in quasars.  For 3C~273 itself, the most important next
step will be to extend our models into the slim disk regime, and possibly
consider nonzero external torques across the innermost stable circular
orbit.  These effects should reduce the discrepancies between the model and
the data in the near ultraviolet and optical portions of the spectrum.

\acknowledgments{We thank Gerard Kriss for generously making available to us
the observational data used in this paper, Beverley Wills for discussions
on possible nonthermal contamination of the big blue bump in 3C~273, and
Tim Sasseen for discussions on far ultraviolet extinction.  This work
was supported by NASA grant NAG5-7075 and NSF grant PHY-9907949.}

\begin{figure}
\figurenum{1}
\plotone{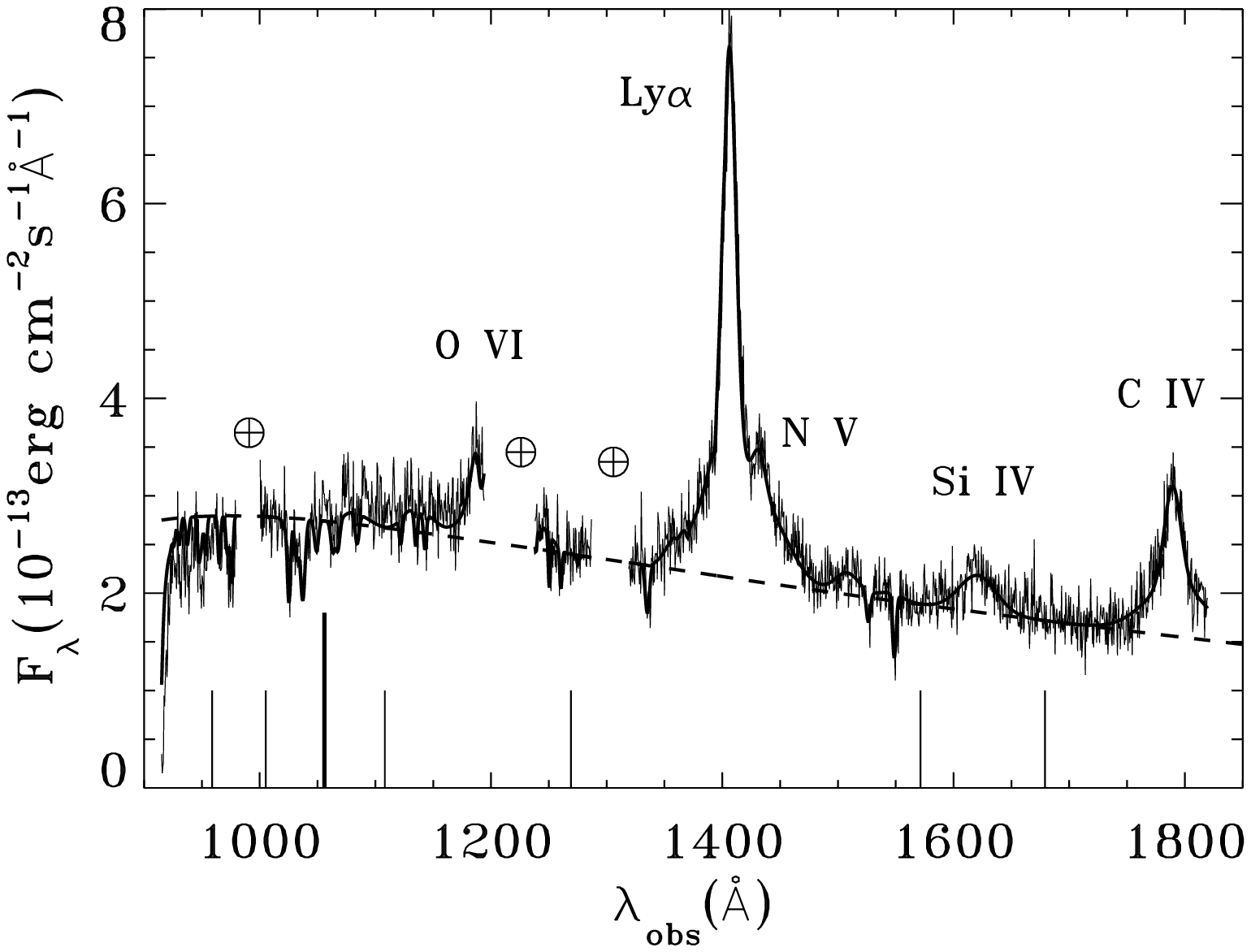}
\vskip 0.1truein
\caption{Best fit non-LTE Kerr disk continuum model to the December 1990 3C~273
{\it HUT} data (noisy curves), assuming a fixed reddening of $E(B-V)=0.02$.
The heavy dashed curve represents the
underlying, reddened continuum model [$M=1.6\times10^{10}$~$\msun$,
$\dot M=4$~$\msun$~yr$^{-1}$, and $\cos i=0.4$], while
the heavy solid curve shows this model with emission and absorption lines
as determined by the fits in KDZL.  Some of the more prominent emission lines
associated with the quasar are marked, and the large vertical line at
$\lambda_{\rm obs}=1056$\AA\/ shows the position of the hydrogen Lyman limit
in the rest frame of the quasar.  Earth symbols indicate wavelength regions
that we have excluded from our fits because of the presence of geocoronal
emission lines.  Short vertical lines denote our chosen continuum points.}
\end{figure}

\begin{figure}
\figurenum{2}
\plotone{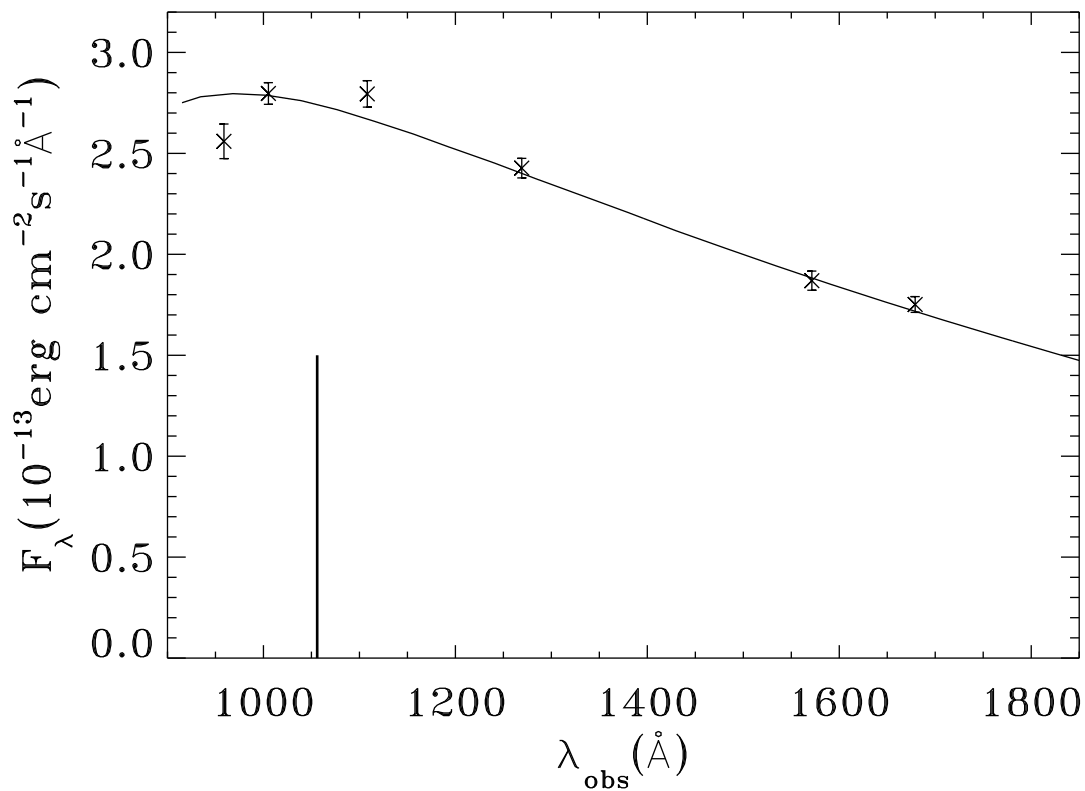}
\vskip 0.1truein
\caption{Best fit non-LTE Kerr disk continuum model to our six chosen continuum
points of the December 1990 {\it HUT} data.  The large vertical line again
shows the position of the hydrogen Lyman limit in the rest frame of the
quasar.  The model parameters are $M=1.6\times10^{10}$~$\msun$,
$\dot M=4$~$\msun$~yr$^{-1}$, $\cos i=0.4$, and $E(B-V)=0.024$.}
\end{figure}

\begin{figure}
\figurenum{3}
\plotone{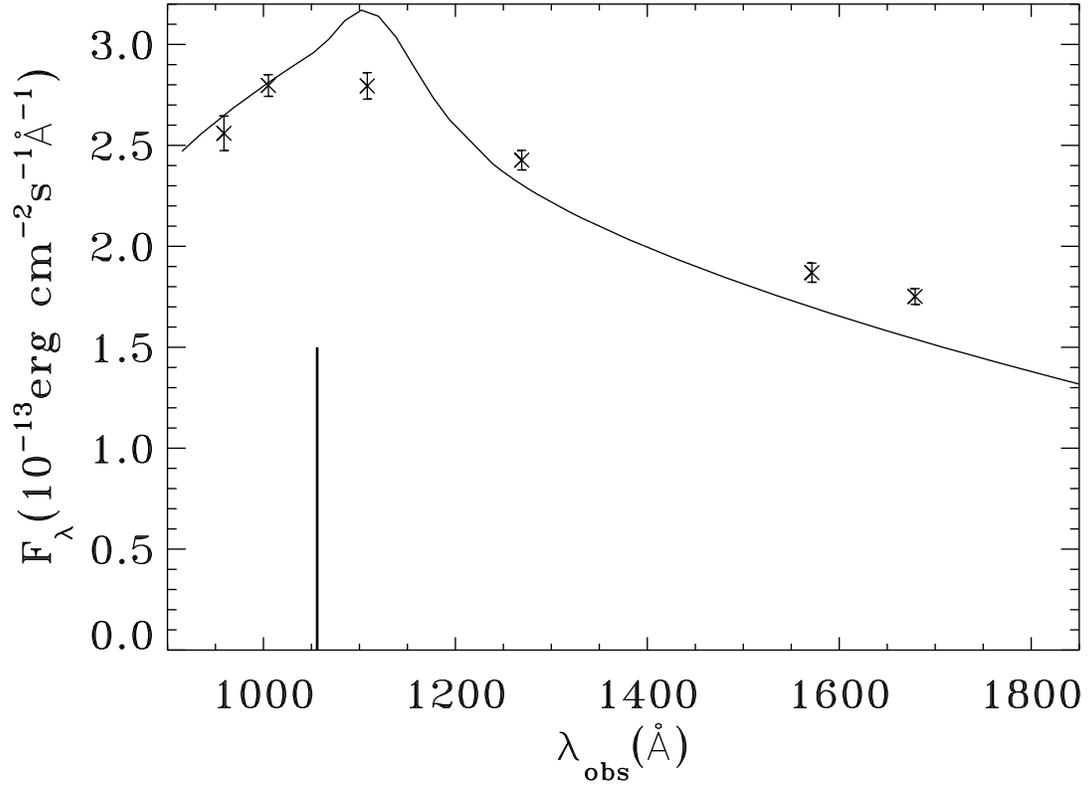}
\vskip 0.1truein
\caption{Same as figure 2, except showing the best fit non-LTE, near face-on
($\cos i=0.99$) Kerr disk continuum model.  The model parameters are
$M=2\times10^9$~$\msun$, $\dot M=4$~$\msun$~yr$^{-1}$, and $E(B-V)=0.0253$.}
\end{figure}

\begin{figure}
\figurenum{4}
\plotone{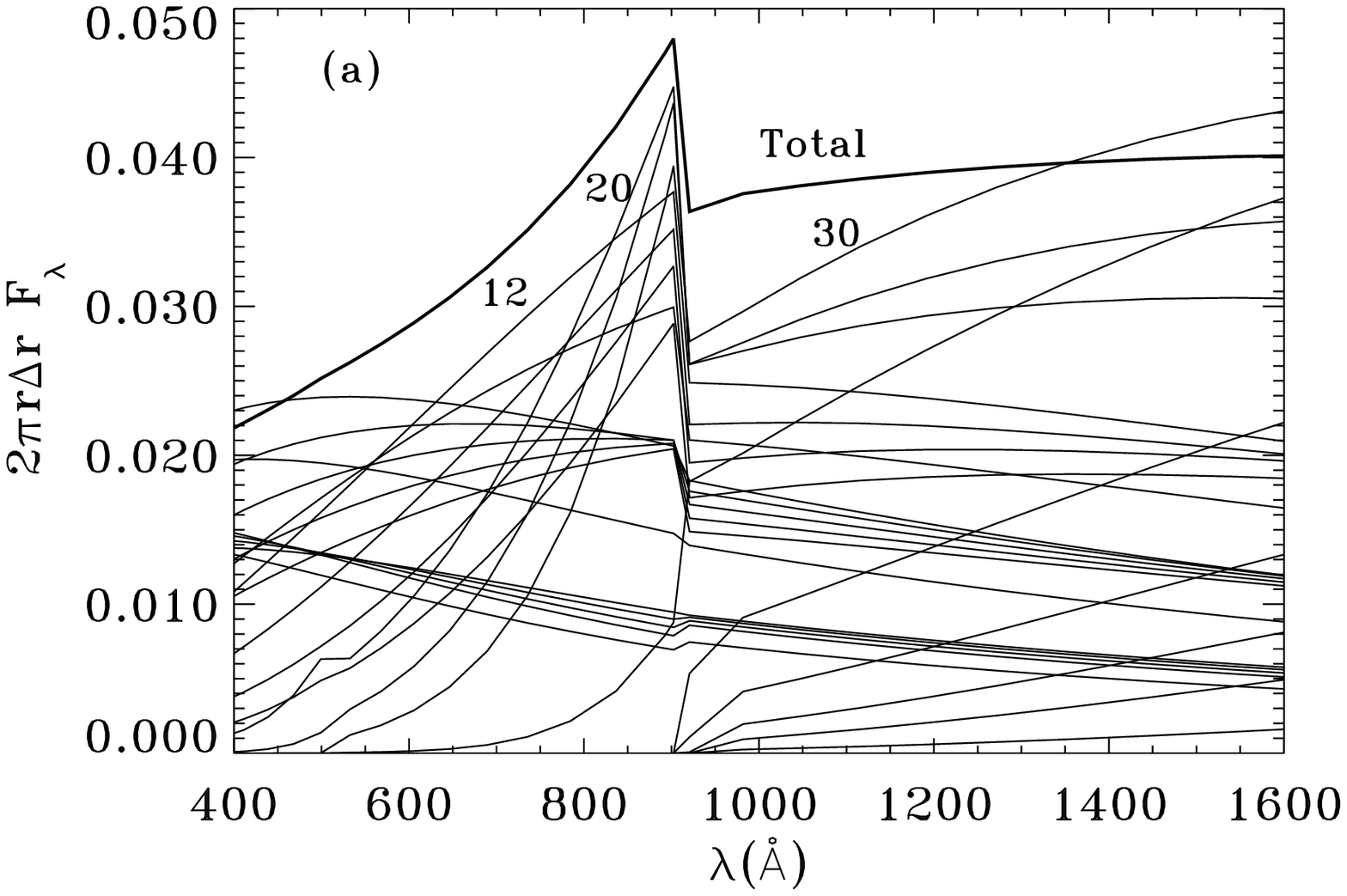}
\end{figure}

\begin{figure}
\figurenum{4}
\plotone{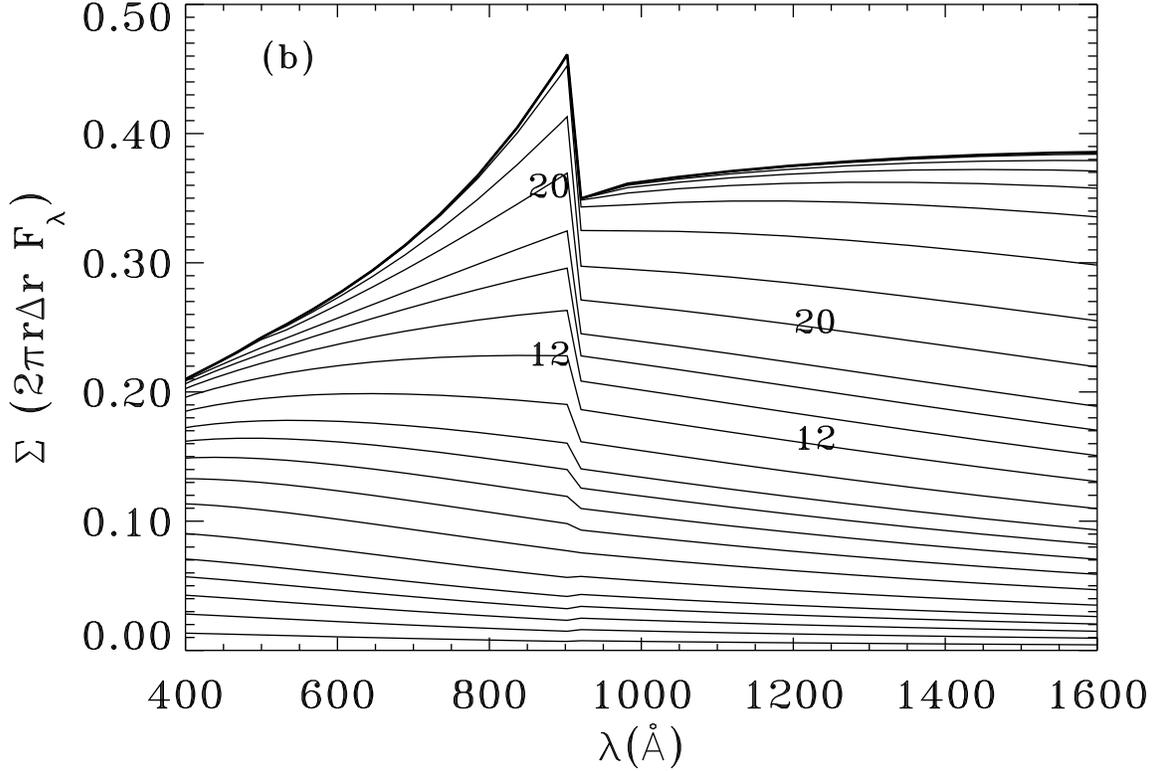}
\vskip 0.1truein
\caption{(a) Local emergent fluxes (in arbitrary units) from the various
annuli contributing to the Kerr disk continuum model in figure 3.  
The flux from each annulus is multiplied by its nominal (Newtonian) surface
area $2\pi r\Delta r$ in order to weight its contribution to the total flux.
The annuli contributing most to the flux around the Lyman limit are labeled
with their radii in units of the gravitational radius $GM/c^2$.
The curve labeled ``Total'' is the average flux from all the
annuli, multiplied by 2.6 for clarity.  (b) Cumulative sum of the local
emergent fluxes.  From bottom to top, the curves represent the contributions
of all annuli out
to radius 1.5, 2, 2.5, 3, 3.5, 4, 5, 6, 7, 8, 9, 10, 12, 14, 16, 18, 20, 25,
30, 40, 50, 60, 70, 80, and 90; in units of the gravitational radius.
The topmost bold curve is the sum.  Convergence is obtained
in the Lyman continuum at about 30 gravitational radii, with annuli
around 12 and 20 gravitational radii providing the dominant contribution.
The cumulative sums out to these two annuli are labeled in the figure.
}
\end{figure}

\begin{figure}
\figurenum{5}
\epsscale{0.85}
\plotone{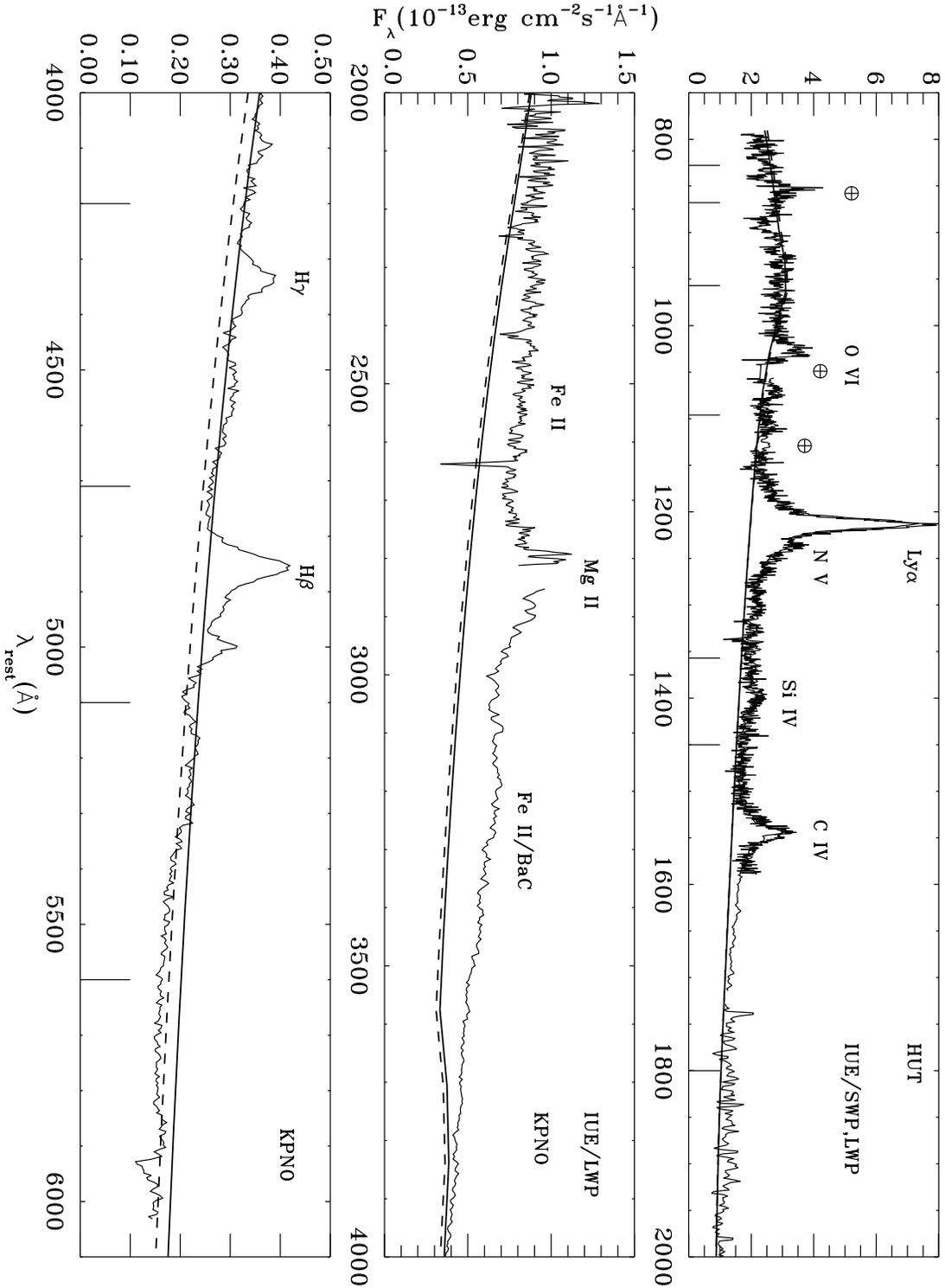}
\vskip 0.1truein
\caption{Best fit near face-on ($\cos i=0.99$) non-LTE Kerr models to the
December 1990 3C~273 multiwavelength data (noisy curves).  The smooth
curves represent the underlying, reddened continuum models including (solid)
and without (dashed) a possible blazar contribution (see text for details).
Prominent emission lines associated with the quasar have been marked
and the various observatories used to acquire data are indicated.  Large
vertical tick marks in the top and bottom figures indicate the locations of
the eleven continuum points used in the fits.}
\end{figure}

\begin{figure}
\figurenum{6}
\plotone{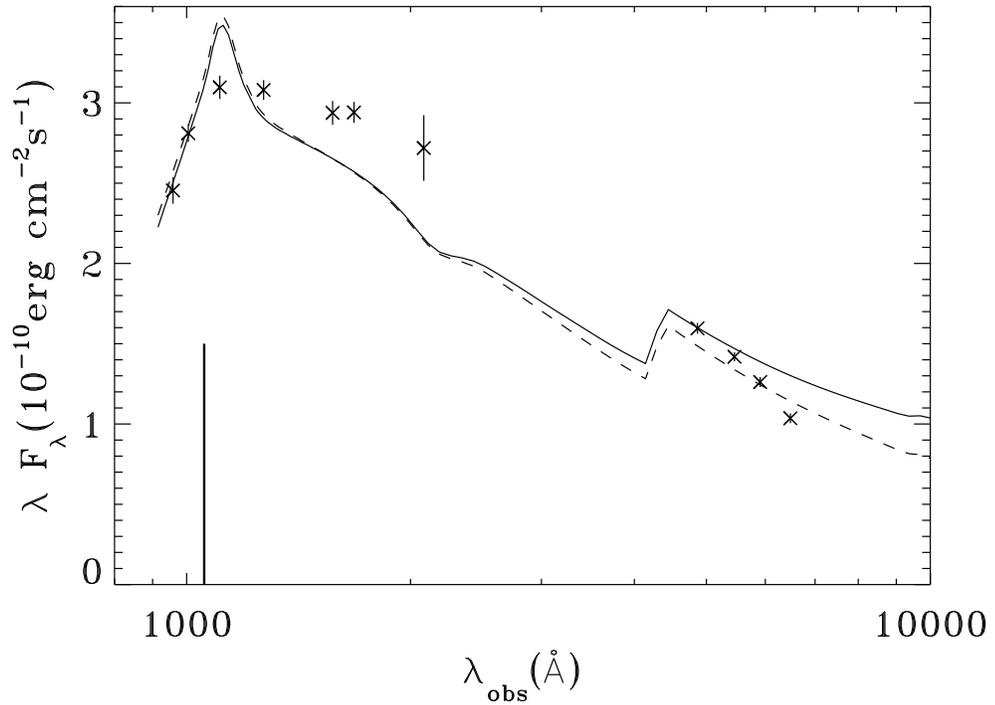}
\vskip 0.1truein
\caption{The same models as shown in figure 5, plotted against the eleven
chosen continuum data points used in the fits.  The vertical line shows the
position of the hydrogen Lyman limit in the rest frame of the quasar.  The
dip in the model spectrum near 2000~\AA\/ is due to the 2200~\AA\/ feature
in the reddening curve.}
\end{figure}

\begin{figure}
\figurenum{7}
\plotone{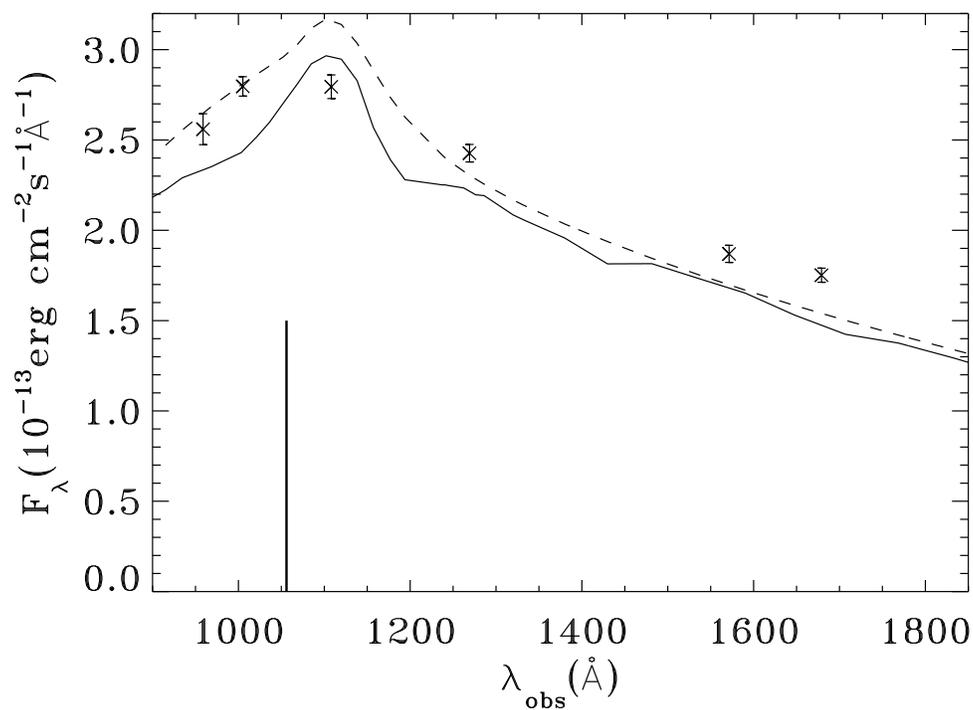}
\vskip 0.1truein
\caption{Same as figure 3, except showing a non-LTE, near face-on
($\cos i=0.99$) Kerr disk continuum model, including the effects of line
blanketing (solid line).  The model parameters are
$M=2\times10^9$~$\msun$, $\dot M=4$~$\msun$~yr$^{-1}$, and $E(B-V)=0.0253$.
The original model of figure 3 without line-blanketing, is shown as the dashed
line.}
\end{figure}

\begin{figure}
\figurenum{8}
\plotone{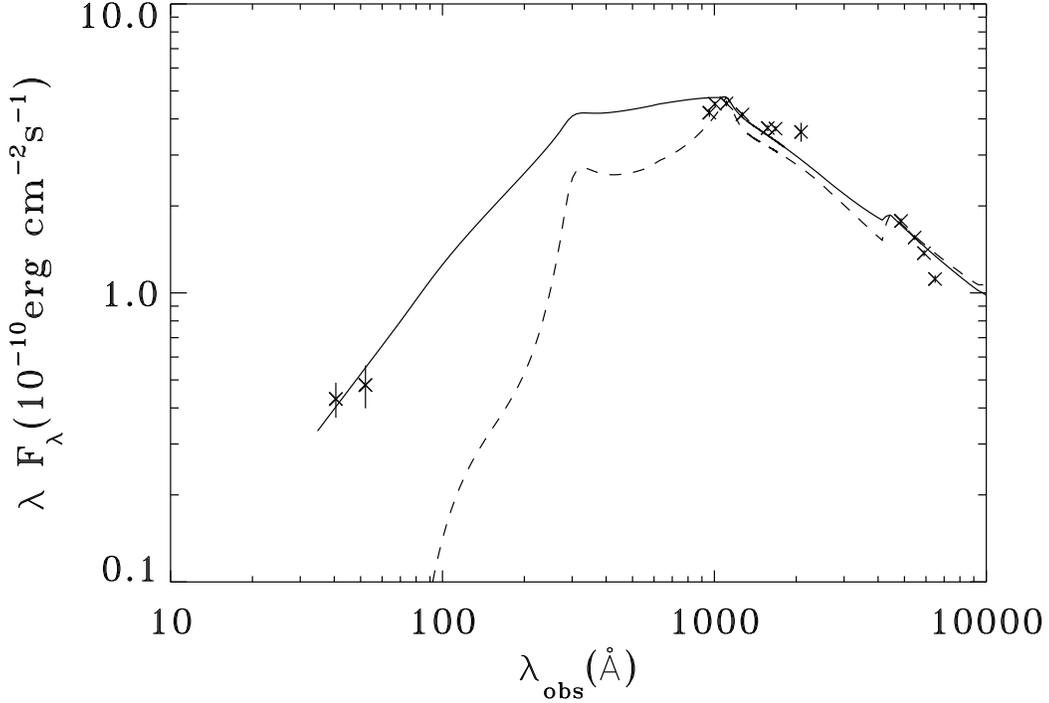}
\vskip 0.1truein
\caption{The energy distribution of 3C~273 from optical to X-rays
during the December 1990 observing epoch.  The two shortest wavelength data
points were measured by {\it ROSAT}, while the longer wavelength points are the
eleven optical/ultraviolet continuum data points used in our spectral fits.
In contrast to all other plots in this paper, the data points here have been
corrected for Galactic hydrogen absorption (in X-rays) and reddening (in the
optical and ultraviolet) with $E(B-V)=0.032$.  The dashed curve shows our best
fit Kerr model from figure 3, with $M=2\times10^9$~$\msun$,
$\dot M=4$~$\msun$~yr$^{-1}$, and $\cos i=0.99$.  The solid curve
shows a Comptonized version of this model designed to go through the {\it ROSAT}
data points.  The Comptonizing medium for this model had a Thomson depth of
unity and an electron temperature of $3.3\times10^8$~K.}
\end{figure}

\begin{figure}
\figurenum{9}
\plotone{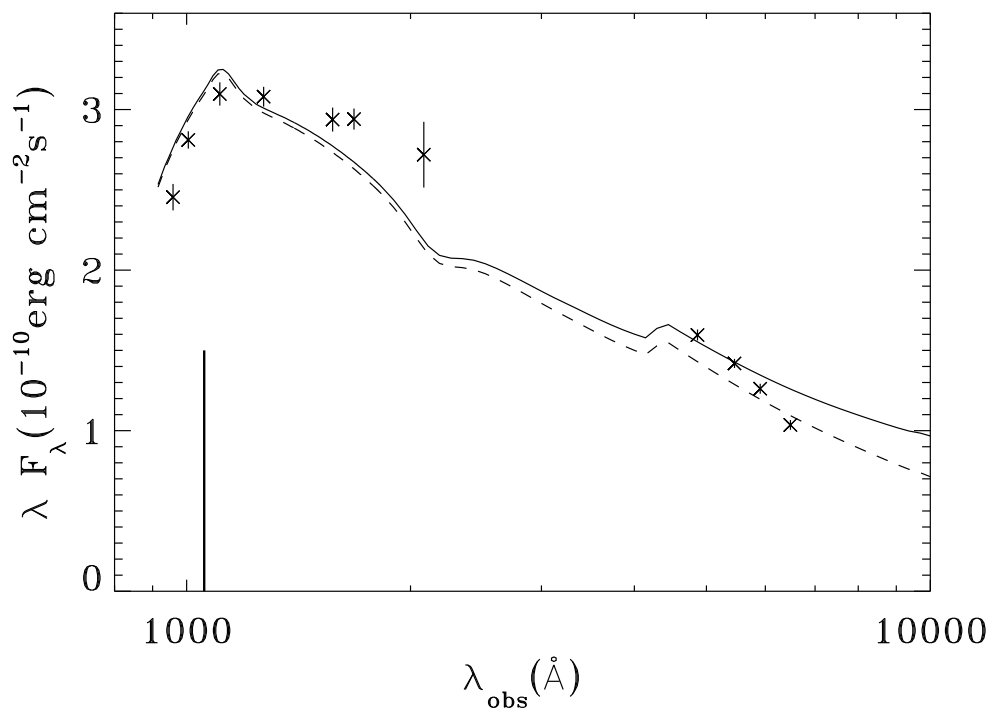}
\vskip 0.1truein
\caption{Same as figure 6 except for our Comptonized near face-on Kerr models.
The model parameters are $M=2\times10^9$~$\msun$, $\dot M=4$~$\msun$~yr$^{-1}$,
$\cos i=0.99$, and $E(B-V)=0.032$.  The Comptonizing medium had a Thomson
depth of unity and an electron temperature of $3.3\times10^8$~K.  The solid
curve has an additional blazar component, while the dashed curve does not have
such a component.
}
\end{figure}

\begin{figure}
\figurenum{10}
\plotone{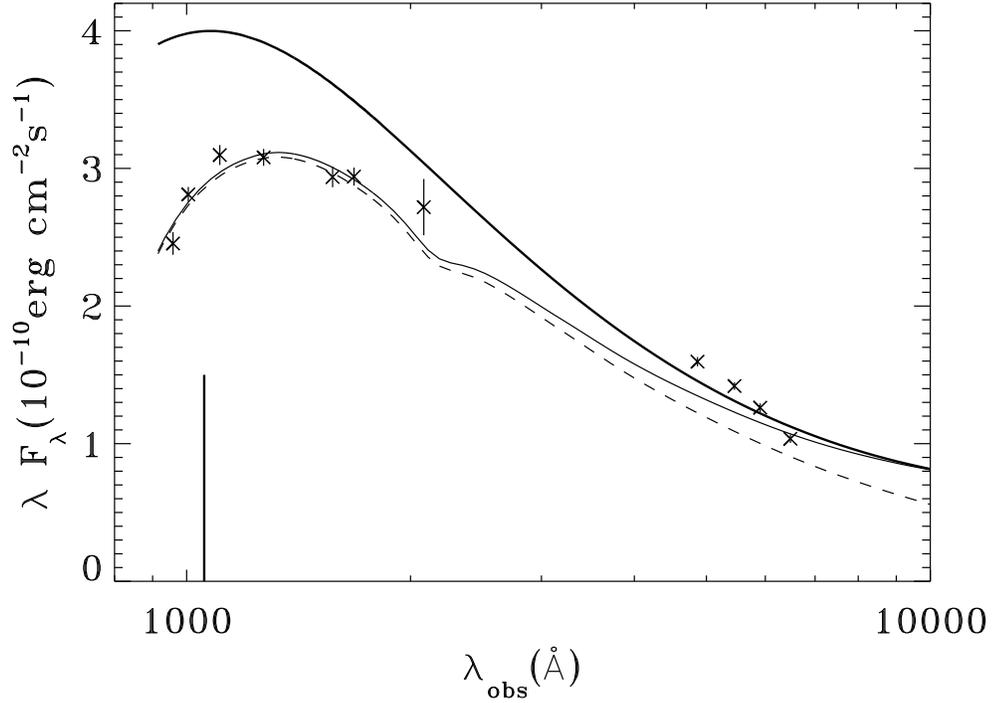}
\vskip 0.1truein
\caption{A multitemperature blackbody, near face-on ($\cos i=0.99$)
disk model which provides a reasonable fit to the far ultraviolet data of
3C~723.  The upper thick solid curve shows the model without reddening
(but with a blazar component), while the lower curves show the reddened model
with (solid) and without (dashed) a blazar component.  The model parameters
are $a/M=0.7$, $M=7\times10^8$~M$_\odot$, $\dot M=6.8$~M$_\odot$~yr$^{-1}$,
and a Galactic reddening of $E(B-V)=0.025$.
}
\end{figure}


\begin{thebibliography}{}

\bibitem[Abraham et al. 1996]{abr96} Abraham, Z., Carrara, E. A., Zensus,
J. A., \& Unwin, S. C. 1996, A\&AS, 115, 543

\bibitem[Abraham \& Romero 1999]{abr99} Abraham, Z., \& Romero, G. E. 1999,
A\&A, 344, 61


\bibitem[Agol 1997]{ago99} Agol, E. 1997, Ph.D. thesis, Univ. California,
Santa Barbara

\bibitem[Agol \& Krolik 2000]{ago00} Agol, E., \& Krolik, J. H. 2000, ApJ, 528,
161

\bibitem[Agol et al. 2001]{ago01} Agol, E., Krolik, J. H., Turner, N., \&
Stone, J. 2001, ApJ, in press

\bibitem[Antonucci et al. 1989]{ant89} Antonucci, R. R. J., Kinney, A. L.,
\& Ford, H. C. 1989, ApJ, 342, 64

\bibitem[Appenzeller et al. 1998]{app98} Appenzeller, I., Krautter, J.,
Mandel, H., Bowyer, S., Dixon, W. V., Hurwitz, M., Barnstedt, J., Grewing,
M., Kappelmann, N., \& Kr\"amer, G. 1998, ApJ, 500, L9

\bibitem[Begelman 2001]{beg01} Begelman, M. C. 2001, ApJ, 551, 897

\bibitem[Boroson \& Green 1992]{bor92} Boroson, T. A., \& Green, R. F. 1992,
ApJS, 80, 109

\bibitem[Cardelli, Clayton, \& Mathis 1989]{car89} Cardelli, J. A., Clayton,
G. C., \& Mathis, J. S. 1989, ApJ, 345, 245

\bibitem[Coleman 1993]{col93} Coleman, H. H. 1993, PhD Thesis, University of
Texas, Austin, TX

\bibitem[Collin 2001]{col01} Collin, S. 2001, in GH Advanced Lectures on the
Starburst-AGN Connection, eds. D. Kunth \& I. Aretxaga, in press
(astro-ph/0101203)

\bibitem[Cunningham 1976]{cun76} Cunningham, C. 1976, ApJ, 208, 534

\bibitem[Cutri et al. 1985]{cut85} Cutri, R. M., Wi\'sniewski, W. Z., Rieke,
G. H., \& Lebofsky, M. J. 1985, ApJ, 296, 423

\bibitem[Czerny \& Pojmanski 1990]{cze90}Czerny, B., \& Pojma\'nski, G. 1990,
MNRAS, 245, 1P

\bibitem[Czerny \& Zbyszewska 1991]{cze91}Czerny, B., \& Zbyszewska, M. 1991,
MNRAS, 249, 634


\bibitem[Francis et al. 1991]{fra91} Francis, P. J., Hewett, P. C., Foltz,
C. B., Chaffee, F. H., Weymann, R. J., \& Morris, S. L. 1991, ApJ, 373, 465

\bibitem[Gammie 1999]{gam99} Gammie, C. F. 1999, ApJ, 522, L57

\bibitem[Hubeny et al. 2000]{hub00} Hubeny, I., Agol, E., Blaes, O., \&
Krolik, J. H. 2000, ApJ, 533, 710 (HABK)

\bibitem[Hubeny et al. 2001]{hub01} Hubeny, I., Blaes, O., Krolik, J. H., \&
Agol, E. 2001, ApJ, in press

\bibitem[Hubeny \& Hubeny 1997]{hub97} Hubeny, I., \& Hubeny, V. 1997,
ApJ, 484, L37


\bibitem[Hubeny \& Hubeny 1998]{hub98} Hubeny, I., \& Hubeny, V. 1998,
in Accretion Processes in Astrophysical Systems: Some Like it Hot!,
ed. S. S. Holt \& T. R. Kallman (Woodbury, NY: AIP), 171

\bibitem[Impey et al. 1989]{imp89} Impey, C. D., Malkan, M. A., \&
Tapia, S. 1989, ApJ, 347, 96

\bibitem[Impey et al. 1995]{imp95} Impey, C., Malkan, M., Webb, W., \& Petry,
C. 1995, ApJ, 440, 80

\bibitem[Kolykhalov \& Sunyaev]{kol84} Kolykhalov, P. I., \& Sunyaev, R. A.
1984, Adv. Space Res., 3, 249

\bibitem[Koratkar 1998]{kor98} Koratkar, A. 1998, in Accretion Processes in
Astrophysical Systems: Some Like it Hot!, ed. S. S. Holt \& T. R. Kallman
(Woodbury, NY: AIP), 150

\bibitem[Koratkar et al. 1995]{kor95} Koratkar, A., Antonucci, R. R. J.,
Goodrich, R. W., Bushouse, H., \& Kinney, A. L. 1995, ApJ, 450, 501

\bibitem[Koratkar \& Blaes 1999]{kor99} Koratkar, A., \& Blaes, O. 1999,
PASP, 111, 1

\bibitem[Koratkar et al. 1992]{kor92} Koratkar, A. P., Kinney, A. L., \&
Bohlin, R. C. 1992, ApJ, 400, 435

\bibitem[Kriss et al. 1999]{kri99} Kriss, G. A., Davidsen, A. F., Zheng, W.,
\& Lee, G. 1999, ApJ, 527, 683 (KDZL)

\bibitem[Krolik 1998]{kro98} Krolik, J. H. 1998, ApJ, 498, L13

\bibitem[Krolik 1999a]{kro99a} Krolik, J. H. 1999a, Active Galactic
Nuclei (Princeton: Princeton Univ. Press)

\bibitem[Krolik 1999b]{kro99b} Krolik, J. H. 1999b, ApJ, 515, L73


\bibitem[Laor 1992]{lao92} Laor, A. 1992, in ``Testing the AGN Paradigm'',
ed. S. S. Holt, S. G. Neff, \& C. M. Urry, 155

\bibitem[Laor et al. 1997]{la97} Laor, A., Jannuzi, B. T., Green, R. F., \&
Boroson, T. A. 1997, ApJ,489, 656

\bibitem[Laor \& Netzer 1989]{ln90} Laor, A., \& Netzer, H. 1989,
MNRAS, 238, 897


\bibitem[Lee et al. 1992]{lee92} Lee, G., Kriss, G. A., \& Davidsen, A. F.
1992, in ``Testing the AGN Paradigm'', ed. S. S. Holt, S. G. Neff, \& C. M.
Urry, 159

\bibitem[Lichti et al. 1995]{lic95} Lichti, G. G., et al. 1995, A\&A, 298, 711

\bibitem[Lockman \& Savage 1995]{loc95} Lockman, F. J., \& Savage, B. D. 1995,
ApJS, 97, 1

\bibitem[Miller \& Stone 2000]{mil00} Miller, K. A., \& Stone, J. M. 2000,
ApJ, 534, 398

\bibitem[Nandra \& Pounds 1994]{nan94} Nandra, K., \& Pounds, K. A. 1994,
MNRAS, 268, 405




\bibitem[Ramos et al. 1997]{ram97} Ramos, E., Kafatos, M., Fruscione, A.,
Bruhweiler, F. C., McHardy, I. M., Hartman, R. C., Titarchuk, L. G., \&
von Montigny, C. 1997, ApJ, 482, 167

\bibitem[Sasseen et al. 2001]{sas01} Sasseen, T. P., Hurwitz, M., Dixon,
W. V., \& Airieau, S. 2001, ApJ, in press

\bibitem[Sembach et al. 2001]{sem01} Sembach, K. R., Howk, J. C., Savage, B. D.,
Shull, J. M., \& Oegerle, W. M. 2001, ApJ, in press

\bibitem[Schlegel, FinkBeiner, \& Davis 1998]{sch98} Schlegel, D. J.,
Finkbeiner, D. P., \& Davis, M. 1998, ApJ, 500, 525
 

\bibitem[Shields 1978]{shi78} Shields, G. A. 1978, Nature, 272, 706

\bibitem[Shields et al. 2000]{shi00} Shields, G. A., Agol, E., \& Blaes, O.
2000, in The Seventh Texas-Mexico Conference on Astrophysics: Flows, Blows
and Glows, ed. W. Lee \& S. Torres-Peimbert, in press

\bibitem[Shields \& Coleman]{shi94} Shields, G. A., \& Coleman, H. H. 1994,
in Theory of Accretion Disks, ed. W. J. Duschl et al. (NATO ASI Ser. C, 417;
Dordrecht: Kluwer), 223



\bibitem[Sincell \& Krolik 1998]{sin98} Sincell, M. W., \& Krolik,
J. H. 1998, ApJ, 496, 737

\bibitem[Storzer et al. 1994]{sto94} St\"orzer, H., Hauschildt, P. H., \&
Allard, F. 1994, ApJ, 437, L91

\bibitem[Sun \& Malkan 1989]{sun89} Sun, W.-H., \& Malkan, M. A. 1989,
ApJ, 346, 68

\bibitem[Szuszkiewicz et al. 1996]{szu96} Szuszkiewicz, E., Malkan, M. A., \&
Abramowicz, M. A. 1996, ApJ, 458, 474

\bibitem[Tanaka et al. 1995]{tan95} Tanaka, Y., et al. 1995, Nature, 375, 659

\bibitem[Turner et al. 2001]{tur01} Turner, N. J., Stone, J. M., \& Sano, T.
2001, ApJ, submitted


\bibitem[Wehrse 1997]{weh97} Wehrse, R. 1997, in ``Accretion Phenomena and
Related Outflows'', ed. D. T. Wickramasinghe, L. Ferrario, \& G. V. Bicknell
(San Francisco: ASP), 162

\bibitem[Wills 1989]{wil89} Wills, B. J. 1989, in ``BL Lac Objects'', ed. L.
Maraschi, T. Maccacaro, \& M.-H. Ulrich (Berlin: Springer), 109


\bibitem[Zheng et al. 1995]{zhe95} Zheng, W., et al. 1995, ApJ, 444, 632

\bibitem[Zheng et al. 1997]{zhe97} Zheng, W., Kriss, G. A., Telfer, R. C.,
Grimes, J. P., \& Davidsen, A. F. 1997, ApJ, 475, 469
 
\end{thebibliography}
\end{document}